\documentclass[pra,twocolumn,floatfix,superscriptaddress,showpacs]{revtex4}
\usepackage{amsmath}

\usepackage{times}
\usepackage{graphicx}

\begin{document}

\title{Dynamical invariants and non-adiabatic geometric phases in open
quantum systems}
\author{M. S. \surname{Sarandy}}
\email{msarandy@if.uff.br}
\affiliation{Departamento de Ci\^encias Exatas, P\'olo Universit\'ario de Volta Redonda,
Universidade Federal Fluminense, Av. dos Trabalhadores 420, Volta Redonda,
27255-125, Rio de Janeiro, Brazil}
\author{E. I. \surname{Duzzioni}}
\email{duzzioni@df.ufscar.br}
\affiliation{Centro de Ci\^encias Naturais e Humanas, Universidade Federal do ABC, R.
Santa Ad\'elia 166, Santo Andr\'e, 09210-170, S\~ao Paulo, Brazil}
\author{M. H. Y. \surname{Moussa}}
\email{miled@ifsc.usp.br}
\affiliation{Instituto de F\'{\i}sica de S\~ao Carlos, Universidade de S\~ao Paulo, Caixa
Postal 369, S\~ao Carlos, 13560-970, S\~ao Paulo, Brazil}

\begin{abstract}
We introduce an operational framework to analyze non-adiabatic Abelian and
non-Abelian, cyclic and non-cyclic, geometric phases in open quantum
systems. In order to remove the adiabaticity condition, we generalize the
theory of dynamical invariants to the context of open systems evolving under
arbitrary convolutionless master equations. Geometric phases are then
defined through the Jordan canonical form of the dynamical invariant
associated with the super-operator that governs the master equation. 
As a by-product, we provide a sufficient condition for the robustness of 
the phase against a given decohering process. We illustrate our results by 
considering a two-level system in a Markovian interaction with
the environment, where we show that the non-adiabatic geometric phase acquired by
the system can be constructed in such a way that it is robust against both 
dephasing and spontaneous emission.
\end{abstract}

\maketitle


\section{Introduction}


Geometric phases (GPs) provide a remarkable mechanism for a quantum system
to keep the memory of its motion as it evolves in Hilbert(-Schmidt)
space. These phase factors depend only on the geometry of the path traversed
by the system during its evolution. In the context of quantum mechanics, GPs
were first obtained by Berry~\cite{Berry:84}, who considered the adiabatic
cyclic evolution of a non-degenerate quantum system isolated from the
contact with a quantum environment. After the seminal work by Berry, the
concept of GPs has been generalized in a number of distinct directions,
e.g., degenerate systems~\cite{Wilczek:84}, non-adiabatic~\cite{Aharonov:87}
and non-clyclic evolutions~\cite{Samuel:88}, etc. Besides its conceptual
importance in quantum mechanics, GPs have also attracted an increasing
attention since their proposal as a tool to achieve fault tolerance in
quantum information processing~\cite{Zanardi:99,Jones:00}.

Motivated by the applications in quantum information, a great effort has
been devoted to analyzing GPs in \textit{open quantum systems}, i.e.,
quantum systems subjected to decoherence due to its interaction with a
quantum environment~\cite{Breuer:Book}. The assumption that a quantum system
is closed is always an idealization and therefore, in order to implement
realistic applications in quantum mechanics, we should be able to estimate the effects of the
surrounding environment on the dynamics of the system. For a number of
physical phenomena, the open system can be conveniently described by a
convolutionless (local) master equation after the degrees of freedom of the
environment are traced out~\cite{Breuer:Book,Alicki:87}. In this context,
several treatments for GPs acquired by the density operator have been
proposed (see, e.g., Refs.~\cite{Ellinas:89,
Gamliel:89,Romero:02,Kamleitner:04,Marzlin:04}). Moreover, in the particular
case of Markovian interaction with the environment, where the system is
described by a master equation in the Lindblad form~\cite{Lindblad:76}, GPs
have also been analyzed through quantum trajectories~\cite%
{Faria:03,Carollo:03, Guridi:05} (see also Ref.~\cite{Bassi:06} for a
further discussion of GPs via stochastic unravelings).

More recently, in the case of adiabatic evolution, Abelian and non-Abelian
GPs in open systems have been generally defined in Ref.~\cite{Sarandy:06}.
This approach was based on an adiabatic approximation previuosly established
for convolutionless master equations~\cite{Sarandy:05} (see also Ref.~\cite%
{Sarandy2:05} for an application of this adiabatic approximation in
adiabatic quantum computation under decoherence and Ref.~\cite{Thunstrom:05}
for an alternative adiabatic approach in weakly coupled open systems).
However, although the adiabatic behavior is usually a very welcome feature
in theoretical models, it can be unsuitable if decoherence times are small.
Therefore, it would be rather desirable to have a general formalism to deal
with non-adiabatic GPs for systems under decoherence. In closed systems, a
useful tool to remove the adiabaticity constraint of geometric phases~\cite%
{Morales:88, Mizrahi:89, Mostafazadeh:98, Mostafazadeh:99} is the theory of
dynamical invariants~\cite{Lewis:69} to treat time-dependent Hamiltonians. 
Indeed, dynamical invariants were recently used in a proposal of an
interferometric experiment to measure non-adiabatic GPs in cavity quantum electrodynamics~%
\cite{Duzzioni:05}.

The aim of this work is to generalize the theory of dynamical invariants to
the context of open quantum systems and to show how this generalization can
be used to establish a general approach for non-adiabatic, Abelian and
non-Abelian, cyclic and non-cyclic, GPs acquired by the components of the density operator
of a system evolving under a convolutionless master equation (see also a
related work in Ref.~\cite{Duzzioni:07}, which introduced a relationship
between GPs and dynamical invariants for a master equation in the Lindblad
form). Within our formalism, we will be able to provide a sufficient condition 
to ensure the robustness of the phase against a given decohering process. 
As an illustration of our result, we will consider a two-level
quantum system (a qubit) interacting with an evironment through a Lindblad
equation. Then, we will find that this system is robust against both
dephasing and spontaneous emission. This generalizes the results for the
robustness against these decohering processes found in the adiabatic case in
Ref.~\cite{Sarandy:05}. Furthermore, in the case of spontaneous emission,
robustness of the non-adiabatic GP is a new feature of our approach, which
should positively impact geometric QC (see, e.g., Ref.~\cite{Guridi:05} for
difficulties in the correction of spontaneous emission).


\section{Dynamical invariants in open systems}


For a closed quantum system, a dynamical invariant $I(t)$ is an Hermitian operator
which satisfies \cite{Lewis:69}%
\begin{equation}
\frac{\partial I}{\partial t}-\frac{1}{i}\left[ H,I\right] =0,
\label{diclosed}
\end{equation}%
where $H$ is the Hamiltonian of the system. Dynamical invariants have
time-independent eigenvalues, implying therefore that their expectation
value is constant, i.e., $d\langle I(t)\rangle /dt=0$.

In order to generalize the concept of a dynamical invariant to the context
of open systems, we consider a general open system described by a
convolutionless master equation 
\begin{equation}
\mathcal{L}\rho =\frac{{\partial {\rho }}}{\partial t},  \label{eq:t-Lind2}
\end{equation}%
where $\rho (t)$ is the density operator, which can be taken as vector in
Hilbert-Schmidt space, and $\mathcal{L}$ is the (usually non-Hermitian)
super-operator which dictates the dynamics of the system. Given an open
system governed by $\mathcal{L}(t)$, we define a dynamical invariant $%
\mathcal{I}(t)$ as a super-operator which satisfies the equation 
\begin{equation}
\frac{\partial \mathcal{I}}{\partial t}-\left[ \mathcal{L},\mathcal{I}\right]
=0.  \label{di}
\end{equation}%
Similarly to the case of closed systems, the eigenvalues of the
super-operator $\mathcal{I}(t)$ will be shown to be time-independent, as
expected for a dynamical invariant. However, note that Eq.~(\ref{di}) does
not uniquely determine $\mathcal{I}(t)$ nor ensures that such a
super-operator exists. The success of our approach will rely therefore on
the possibility of constructing non-trivial (time-dependent) dynamical
invariants, which can fortunately be found in a number of interesting
examples.

The super-operator $\mathcal{I}(t)$ is in general non-Hermitian, which means
that it will not always exhibit a basis of eigenstates. However we can
construct a right basis $\{|\mathcal{D}_{\alpha }^{(i)}\rangle \rangle \}$
and a left basis $\{\langle \langle \mathcal{E}_{\alpha }^{(i)}|\}$ in
Hilbert-Schmidt space based on the Jordan canonical form of $\mathcal{I}(t)$~%
\cite{Horn:book}. Here, the double-ket notation is used to emphasize that 
these vectors are defined in the space state of linear operators instead of the 
ordinary Hilbert space. This construction is analogous to the procedure developed
in Ref.~\cite{Sarandy:05}, but using now the Jordan decomposition of $%
\mathcal{I}(t)$ rather than $\mathcal{L}(t)$. It can be shown (see Ref.~\cite%
{Sarandy:05} or Appendix A of Ref.~\cite{Sarandy:06}) that left and right
basis vectors can always be chosen such that they have the properties 
\begin{eqnarray}
\mathcal{I}\,|\mathcal{D}_{\alpha }^{(i)}\rangle \rangle &=&\lambda _{\alpha
}|\mathcal{D}_{\alpha }^{(i)}\rangle \rangle +|\mathcal{D}_{\alpha
}^{(i-1)}\rangle \rangle \,,  \label{rightbasis} \\
\langle \langle \mathcal{E}_{\alpha }^{(i)}|\,\mathcal{I} &=&\langle \langle 
\mathcal{E}_{\alpha }^{(i)}|\lambda _{\alpha }+\langle \langle \mathcal{E}%
_{\alpha }^{(i+1)}|\,,  \label{leftbasis}
\end{eqnarray}%
where $|\mathcal{D}_{\alpha }^{(-1)}\rangle \rangle \equiv 0$ and $\langle
\langle \mathcal{E}_{\alpha }^{(n_{\alpha })}|\equiv 0$, with the index $%
\alpha $ enumerating each Jordan block and the index $i$ enumerating the
basis vectors inside each Jordan block, with $i=0,...,n_{\alpha }-1$ ($%
n_{\alpha }$ is the dimension of the block $\alpha $). Moreover, left and
right vectors satisfy the orthonormality condition 
\begin{equation}
\langle \langle \mathcal{E}_{\alpha }^{(i)}|\mathcal{D}_{\beta
}^{(j)}\rangle \rangle =\delta _{\alpha \beta }\delta ^{ij}\,.
\label{oc}
\end{equation}%
The eigenvalues of $\mathcal{I}(t)$ are denoted by $\lambda _{\alpha }$ and
the left and right eigenvectors of $\mathcal{I}(t)$ are denoted by $|%
\mathcal{D}_{\alpha }^{(0)}\rangle \rangle $ and $\langle \langle \mathcal{E}%
_{\alpha }^{(n_{\alpha }-1)}|$, respectively. Taking the derivative of Eq.~(%
\ref{rightbasis}) with respect to time (denoted by the dot symbol below), we
obtain 
\begin{equation}
\dot{\mathcal{I}}|\mathcal{D}_{\alpha }^{(i)}\rangle \rangle +\mathcal{I}|%
\dot{\mathcal{D}}_{\alpha }^{(i)}\rangle \rangle =\dot{\lambda _{\alpha }}|%
\mathcal{D}_{\alpha }^{(i)}\rangle \rangle +\lambda _{\alpha }|\dot{\mathcal{%
D}}_{\alpha }^{(i)}\rangle \rangle +|\dot{\mathcal{D}}_{\alpha
}^{(i-1)}\rangle \rangle \,.  \label{step1}
\end{equation}%
Projection of Eq.~(\ref{step1}) in $\langle \langle \mathcal{E}_{\beta
}^{(j)}|$ yields 
\begin{eqnarray}
&&\langle \langle \mathcal{E}_{\beta }^{(j)}|\dot{\mathcal{I}}|\mathcal{D}%
_{\alpha }^{(i)}\rangle \rangle =\dot{\lambda}_{\alpha }\delta _{\alpha
\beta }\delta ^{ij}+\left( \lambda _{\alpha }-\lambda _{\beta }\right)
\langle \langle \mathcal{E}_{\beta }^{(j)}|\dot{\mathcal{D}}_{\alpha
}^{(i)}\rangle \rangle \,  \nonumber \\
&&+\langle \langle \mathcal{E}_{\beta }^{(j)}|\dot{\mathcal{D}}_{\alpha
}^{(i-1)}\rangle \rangle -\langle \langle \mathcal{E}_{\beta }^{(j+1)}|\dot{%
\mathcal{D}}_{\alpha }^{(i)}\rangle \rangle \,.  \label{projection}
\end{eqnarray}%
On the other hand, from the definition of a dynamical invariant, given by
Eq.~(\ref{di}), we get 
\begin{eqnarray}
&&\langle \langle \mathcal{E}_{\beta }^{(j)}|\dot{\mathcal{I}}|\mathcal{D}%
_{\alpha }^{(i)}\rangle \rangle =\left( \lambda _{\alpha }-\lambda _{\beta
}\right) \langle \langle \mathcal{E}_{\beta }^{(j)}|\mathcal{L}|\mathcal{D}%
_{\alpha }^{(i)}\rangle \rangle \hspace{1.3cm}  \nonumber \\
&&+\langle \langle \mathcal{E}_{\beta }^{(j)}|\mathcal{L}|\mathcal{D}%
_{\alpha }^{(i-1)}\rangle \rangle -\langle \langle \mathcal{E}_{\beta
}^{(j+1)}|\mathcal{L}|\mathcal{D}_{\alpha }^{(i)}\rangle \rangle \,.
\label{di2}
\end{eqnarray}%
By inserting Eq.~(\ref{di2}) into Eq.~(\ref{projection}), we obtain 
\begin{eqnarray}
&&\dot{\lambda}_{\alpha }\delta _{\alpha \beta }\delta ^{ij}=\left( \lambda
_{\alpha }-\lambda _{\beta }\right) \langle \langle \mathcal{E}_{\beta
}^{(j)}|\mathcal{O}|\mathcal{D}_{\alpha }^{(i)}\rangle \rangle  \nonumber \\
&&+\langle \langle \mathcal{E}_{\beta }^{(j)}|\mathcal{O}|\mathcal{D}%
_{\alpha }^{(i-1)}\rangle \rangle -\langle \langle \mathcal{E}_{\beta
}^{(j+1)}|\mathcal{O}|\mathcal{D}_{\alpha }^{(i)}\rangle \rangle
\label{finaldi}
\end{eqnarray}%
where 
\begin{equation}
\mathcal{O}\equiv \mathcal{L}-\frac{\partial }{\partial t}\,.
\end{equation}%
Let us assume, from now on, that $n_{\alpha }=1$, i.e., the Jordan blocks
are one-dimensional (1D). This means that we are assuming that we were able
to find a diagonalizable $\mathcal{I}(t)$ (even though it can be
non-Hermitian). As we will show below, Abelian GPs will be associated with
the situation where $\mathcal{I}(t)$ has non-degenerate 1D Jordan blocks
while non-Abelian phases will be associated with the situation where $%
\mathcal{I}(t)$ displays degenerate 1D Jordan blocks. For multi-dimensional
Jordan blocks, we should proceed by a case by case analysis, with no general
treatment available.

Therefore, assuming 1D Jordan blocks, we have 
\begin{equation}
\dot{\lambda}_{\alpha }\delta _{\alpha \beta }\delta ^{ij}=\left( \lambda
_{\alpha }-\lambda _{\beta }\right) \langle \langle \mathcal{E}_{\beta
}^{(j)}|\mathcal{O}|\mathcal{D}_{\alpha }^{(i)}\rangle \rangle \,,
\end{equation}%
where, now, the indices $i$ and $j$ appearing in  both 
$\{|\mathcal{D}_{\alpha }^{(i)}\rangle \rangle \}$ and $\{\langle \langle \mathcal{E}_{\alpha }^{(j)}|\}$ 
account for degenerate states, namely, states such that $\lambda _{\alpha }=\lambda _{\beta}$, whichever 
$\alpha$ and $\beta$. Observe that for $\alpha =\beta $ and $i=j$, we obtain $\dot{\lambda}_{\alpha }=0$, 
which implies that the dynamical invariant has indeed time-independent
eigenvalues. Moreover, taking indices $\alpha $ and $\beta $ such that $%
\lambda _{\alpha }\neq \lambda _{\beta }$, we obtain 
\begin{equation}
\langle \langle \mathcal{E}_{\beta }^{(j)}|\mathcal{O}|\mathcal{D}_{\alpha
}^{(i)}\rangle \rangle =0\,\,\,\,\,\,\,(\lambda _{\alpha }\neq \lambda
_{\beta })\,.  \label{c2}
\end{equation}%
Eq.~(\ref{c2}) provides the fundamental condition that will allow for the
definition of non-adiabatic GPs.


\section{Non-adiabatic GPs via dynamical invariants}



\subsection{Abelian case}


\label{ABcase}

Let us assume that the eigenvalues of $\mathcal{I}(t)$ are non-degenerate,
i.e., $\lambda _{\alpha }=\lambda _{\beta }\Rightarrow \alpha =\beta $. In
order to simplify the notation, we will omit the upper index $i$ of the
right and left vectors in the Abelian case. Let us take the density operator 
$\rho $ as a vector in Hilbert-Schmidt space and expand it in the right
basis $\{|\mathcal{D}_{\alpha }\rangle \rangle \}$ 
\begin{equation}
|\rho \rangle \rangle =\sum_{\alpha }c_{\alpha }|\mathcal{D}_{\alpha
}\rangle \rangle   \label{rho}
\end{equation}%
By inserting Eq.~(\ref{rho}) into the master equation~(\ref{eq:t-Lind2}) and
projecting it in $\langle \langle \mathcal{E}_{\beta }|$, we obtain 
\begin{equation}
\dot{c}_{\beta }=\sum_{\alpha }c_{\alpha }\langle \langle \mathcal{E}_{\beta
}|\mathcal{O}|\mathcal{D}_{\alpha }\rangle \rangle \,.  \label{rhocoeff}
\end{equation}%
By using Eq.~(\ref{c2}), we can get rid of the sum in Eq.~(\ref{rhocoeff}),
which implies 
\begin{equation}
\dot{c}_{\beta }=c_{\beta }\langle \langle \mathcal{E}_{\beta }|\mathcal{O}|%
\mathcal{D}_{\beta }\rangle \rangle \,.  \label{rhocoeff2}
\end{equation}%
Solving Eq.~(\ref{rhocoeff2}), we obtain 
\begin{equation}
c_{\beta }(t)=c_{\beta }(0)e^{-\int_{0}^{t}\langle \langle \mathcal{E}%
_{\beta }|\frac{\partial }{\partial t^{\prime }}|\mathcal{D}_{\beta }\rangle
\rangle dt^{\prime }}e^{\int_{0}^{t}\langle \langle \mathcal{E}_{\beta }|%
\mathcal{L}|\mathcal{D}_{\beta }\rangle \rangle dt^{\prime }}
\label{rhocoeff3}
\end{equation}%
Therefore, each right eigenvector $|\mathcal{D}_{\beta }\rangle \rangle $ in
the expansion of $\rho $ gets multiplied by a phase. The first exponential
in Eq.~(\ref{rhocoeff3}) gives origin to the geometric contribution of the
phase whereas the second exponential generates the dynamical sector. 
The geometric phase must be gauge invariant, i.e. it cannot be modified (or eliminated)  
by a multiplication of the basis vectors 
$\{|\mathcal{D}_{\alpha}\rangle \rangle \}$ or $\{\langle \langle \mathcal{E}_{\alpha}|\}$ 
by a local (time-dependent) complex factor. Indeed, let us consider the 
redefinition 
$|\mathcal{D}^{\prime}_{\alpha}\rangle \rangle = \chi(t)e^{i\nu (t)} |\mathcal{D}_{\alpha}\rangle \rangle$ 
($\chi(t)\ne 0 \,\,$, $\forall t$). 
For the left vectors, the orthonormality condition, given by Eq.~(\ref{oc}), imposes that 
$\langle \langle \mathcal{E}^{\prime}_{\alpha}|=\langle \langle \mathcal{E}_{\alpha}|\chi^{-1}e^{-i\nu (t)}$. 
Gauge invariance under these transformations for an arbitrary (cyclic or non-cyclic) path in Hilbert-Schmidt 
space is  achieved by adding a new term in the expression of the GP in Eq. (\ref{rhocoeff3}), resulting in
\begin{equation}
\varphi _{\beta }=\ln \left( \langle \langle \mathcal{E}_{\beta }(0)|%
\mathcal{D}_{\beta }(t)\rangle \rangle \right) -\int_{0}^{t}\langle \langle 
\mathcal{E}_{\beta }(t^{\prime })|\frac{\partial }{\partial t^{\prime }}|%
\mathcal{D}_{\beta }(t^{\prime })\rangle \rangle dt^{\prime }.
\label{opgp}
\end{equation}
By a direct inspection, it can be shown that $\varphi_{\beta }$ is gauge invariant. This is 
analogous to the procedure used in Ref.~\cite{Samuel:88} to extend Berry phases for non-cyclic paths 
in closed systems. 
The contribution coming from the term 
$\ln \left( \langle \langle \mathcal{E}_{\beta }(0)
|\mathcal{D}_{\beta }(t)\rangle \rangle \right)$ 
may affect the visibility of the phase, since $\langle \langle \mathcal{E}_{\beta }(0)|\mathcal{D}_{\beta}(t)\rangle \rangle$ 
is not necessarily a complex number with modulus 1. 
Moreover, note that for a cyclic path of the basis vectors, i.e, 
$|\mathcal{D}_{\alpha}(t)\rangle \rangle = |\mathcal{D}_{\alpha}(0)\rangle \rangle$,  we have 
$\ln \left( \langle \langle \mathcal{E}_{\beta }(0)|%
\mathcal{D}_{\beta }(0)\rangle \rangle \right) = \ln 1 = 0$. Therefore, for the cyclic GP, no extra term 
should be added, with $\varphi_{\beta}$ simplifying to 
\begin{equation}
\varphi _{\beta }^{\textrm{cyclic}}= -\int_{0}^{t}\langle \langle 
\mathcal{E}_{\beta }(t^{\prime })|\frac{\partial }{\partial t^{\prime }}|%
\mathcal{D}_{\beta }(t^{\prime })\rangle \rangle dt^{\prime}\,\,\, (\textrm{cyclic}\,\,\, \textrm{path}).
\label{cpgp}
\end{equation}
Observe also that the phases defined above are non-adiabatic, since no adiabaticity 
requirement has been imposed in any step of our derivation.


\subsection{Non-Abelian case}


Let us consider now the case of 1D degenerate Jordan blocks and expand the
density operator as 
\begin{equation}
|\rho \rangle \rangle =\sum_{\alpha =1}^{m}\sum_{j}c_{\alpha }^{(j)}|%
\mathcal{D}_{\alpha }^{(j)}\rangle \rangle ,  \label{rhoexpdeg}
\end{equation}%
where $m$ is the number of Jordan blocks and $j$ identifies all the right
eigenvectors $|\mathcal{D}_{\alpha }^{(j)}\rangle \rangle $ of $\mathcal{I}%
(t)$ associated with the eigenvalue $\lambda _{\alpha }$. Similarly as in
the non-degenerate case, we insert Eq.~(\ref{rhoexpdeg}) into Eq.~(\ref%
{eq:t-Lind2}) and project the result in $\langle \langle \mathcal{E}_{\beta
}^{(i)}|$, yielding 
\begin{equation}
\dot{c}_{\beta }^{(i)}=\sum_{\alpha =1}^{m}\sum_{j=1}c_{\alpha
}^{(j)}\langle \langle \mathcal{E}_{\beta }^{(i)}|\mathcal{O}|\mathcal{D}%
_{\alpha }^{(j)}\rangle \rangle \,.  \label{rhocoeff4}
\end{equation}%
By making use of Eq.~(\ref{c2}), we obtain 
\begin{equation}
\dot{c}_{\beta }^{(i)}=\sum_{j=1}c_{\beta }^{(j)}\langle \langle \mathcal{E}%
_{\beta }^{(i)}|\mathcal{O}|\mathcal{D}_{\beta }^{(j)}\rangle \rangle \,.
\label{rhocoeff5}
\end{equation}%
Now let us define the matrix $M_{\beta }$, whose elements are given by 
\begin{equation}
M_{\beta }^{(ij)}=\langle \langle \mathcal{E}_{\beta }^{(i)}|\mathcal{O}|%
\mathcal{D}_{\beta }^{(j)}\rangle \rangle =H_{\beta }^{(ij)}+A_{\beta
}^{(ij)}\,,
\end{equation}%
with 
\begin{eqnarray}
H_{\beta }^{(ij)} &=&\langle \langle \mathcal{E}_{\beta }^{(i)}|\mathcal{L}|%
\mathcal{D}_{\beta }^{(j)}\rangle \rangle ,  \nonumber \\
A_{\beta }^{(ij)} &=&-\langle \langle \mathcal{E}_{\beta }^{(i)}|\frac{%
\partial }{\partial t}|\mathcal{D}_{\beta }^{(j)}\rangle \rangle \,.
\end{eqnarray}%
Note that $H_{\beta }$ plays the role of a non-Abelian dynamical phase while 
$A_{\beta }$ will correspond to a geometrical contribution to the total
phase. Moreover, by defining the vector $\overrightarrow{\mathbf{c}}_{\beta }=(c_{\beta
}^{1},\ldots c_{\beta }^{N})^{t}$ in Hilbert-Schmidt space, with $N$
denoting the degree of degeneracy, we get 
\begin{equation}
{\overrightarrow{\dot{\mathbf{c}}}}_{\beta }=M_{\beta }\,%
\overrightarrow{\mathbf{c}}_{\beta }\,,
\end{equation}%
whose formal solution is 
\begin{equation}
\overrightarrow{\mathbf{c}}_{\beta }(t)=\mathcal{U}_{\beta }\,%
\overrightarrow{\mathbf{c}}_{\beta }(0)\,,
\end{equation}%
with 
\begin{equation}
\mathcal{U}_{\beta }=\mathcal{T}\exp \left[ \int_{0}^{t}\left[ H_{\beta
}(t^{\prime })+A_{\beta }(t^{\prime })\right] dt^{\prime }\right] ,
\end{equation}%
where $\mathcal{T}$ is the time-ordering operator. It is important to note
that the matrices $H_{\beta }$ and $A_{\beta }$ do not commute in general.
This means that, in the non-Abelian case, the dynamical and GPs may not be
easily splitted up. This is indeed a feature which also appears in
closed systems for non-adiabatic non-Abelian phases~\cite{Mostafazadeh:98,
Mostafazadeh:99, Anandan:88}.
By assuming that $\left[\int A_\beta(t)dt\, , \, \int H_\beta\right(t)dt]=0$, 
we can extend the the reasoning in Ref. \cite{Duzzioni:07} for the
Hilber-Schmidt space, with the noncyclic non-Abelian GP getting the form
\[
\exp \Phi _{\beta }=\mathcal{W}_{\beta }(t)\mathcal{T}\exp \left[
\int_{0}^{t}A_{\beta }(t^{\prime })dt^{\prime }\right] ,
\]%
where $\mathcal{W}_{\beta }$ is the overlap matrix, whose 
elements are given by $\mathcal{W}_{\beta }^{(ij)}(t)=\langle \langle \mathcal{E}_{\beta
}^{(i)}(0)|\mathcal{D}_{\beta }^{(j)}(t)\rangle \rangle$. The presence of the overlap matrix 
ensures the gauge invariance of the non-Abelian GP, which can be verified by a similar inspection as 
that discussed in Subsection~\ref{ABcase}. Moreover, note that $\mathcal{W}_{\beta }$ reduces to the 
identity for cyclic evolutions.


\subsection{Adiabatic limit}


Let us turn now to an observation about the adiabatic regime. The GPs
defined in the previous sections will get reduced to the adiabatic case
introduced in Ref.~\cite{Sarandy:06} for the choice of a slowly varying
dynamical invariant. Indeed, let us suppose that 
\begin{equation}
\frac{\partial \mathcal{I}}{\partial t}\approx 0\,.  \label{Inv}
\end{equation}%
In this case, by taking into account Eq.~(\ref{di}), we will obtain $\left[ 
\mathcal{L},\mathcal{I}\right] \approx 0$. Then, by assuming that both $%
\mathcal{L}$ and $\mathcal{I}$ are diagonalizable, it follows that they will
have a common basis of eigenstates. Therefore, under the condition (\ref{Inv}), 
the non-adiabatic basis, given by eigenstates of $\mathcal{I}$
will exactly be the same as the adiabatic basis, given by the eigenstates of 
$\mathcal{L}$.


\section{Non-adiabatic GP for a two-level system under decoherence}


Let us examine the GP acquired by a two-level system described by the free
Hamiltonian 
\begin{equation}
H=\frac{\omega }{2}\sigma _{z}.  \label{Htls}
\end{equation}%
Under decoherence in a Markovian environment, the dynamics of the system
will be governed by the Lindblad equation~\cite{Lindblad:76} 
\begin{equation}
\frac{{\partial {\rho }}}{\partial t}=-i\left[ H,\rho \right] -\frac{1}{2}%
\sum_{i}\left( \Gamma _{i}^{\dagger }\Gamma _{i}\rho +\rho \Gamma
_{i}^{\dagger }\Gamma _{i}-2\Gamma _{i}\rho \Gamma _{i}^{\dagger }\right) .
\label{eq:t-Lind3}
\end{equation}


\subsection{Robustness under dephasing}


\label{deph}

Let us start by taking the case of dephasing, where $\Gamma (t)=\gamma
_{d}\sigma _{z}$. In this case, the super-operator $\mathcal{L}$ can be
written as (see Appendix~\ref{A}) 
\begin{equation}
\mathcal{L}=\left( 
\begin{array}{cccc}
0 & 0 & 0 & 0 \\ 
0 & -2\gamma _{d}^{2} & -\omega  & 0 \\ 
0 & \omega  & -2\gamma _{d}^{2} & 0 \\ 
0 & 0 & 0 & 0%
\end{array}%
\right)   \label{Ldp}
\end{equation}%
Therefore, $\mathcal{L}$ has a $2\times 2$ matrix representation given by 
\begin{equation}
\mathcal{L}=\left( 
\begin{array}{cc}
-2\gamma _{d}^{2} & -\omega  \\ 
\omega  & -2\gamma _{d}^{2}%
\end{array}%
\right) \,.  \label{Ldp2}
\end{equation}%
Let us look for a simple non-trivial super-operator $\mathcal{I}(t)$, which
we propose to take the form 
\begin{equation}
\mathcal{I}=\left( 
\begin{array}{cc}
\alpha (t) & \beta (t) \\ 
\gamma (t) & \delta (t)%
\end{array}%
\right) \,,  \label{Idp}
\end{equation}%
where $\alpha (t)$, $\beta (t)$, $\gamma (t)$, and $\delta (t)$ are
time-dependent well-behaved functions. Now, it follows an important fact
about the robustness of the non-adiabatic GP. For arbitrary time-dependent
functions $\alpha (t)$, $\beta (t)$, $\gamma (t)$, and $\delta (t)$, we have
that the commutator $\left[ \mathcal{L},\mathcal{I}\right] $ is independent
of the dephasing parameter $\gamma _{d}$. Indeed 
\begin{equation}
\left[ \mathcal{L},\mathcal{I}\right] =\omega \left( 
\begin{array}{cc}
-\beta -\gamma  & \alpha -\delta  \\ 
\alpha -\delta  & \beta +\gamma 
\end{array}%
\right) \,.  \label{LIComm}
\end{equation}%
Due to this property, we can construct a non-trivial (time-dependent)
super-operator $\mathcal{I}(t)$ that is independent of $\gamma _{d}$. This
operator will generate right and left bases which are also independent of $%
\gamma _{d}$. Hence the GP acquired by the density operator $\rho $ will
keep the independence of $\gamma _{d}$, exhibitng therefore robustness
against dephasing.

Let us analyze in details the GP acquired during a cyclic path of the left
and right vectors. By imposing Eq.~(\ref{di}), we will get a set of coupled
differential equations 
\begin{eqnarray}
\dot{\alpha} &=&-\omega \left( \beta +\gamma \right) \,,  \nonumber \\
\dot{\beta} &=&\omega \left( \alpha -\delta \right) \,,  \nonumber \\
\dot{\gamma} &=&\omega \left( \alpha -\delta \right) \,,  \nonumber \\
\dot{\delta} &=&\omega \left( \beta +\gamma \right) \,.  \label{diffeqs}
\end{eqnarray}%
The solution of this set of equations yields 
\begin{equation}
\mathcal{I}=\left( 
\begin{array}{cc}
\alpha (t) & \beta (t) \\ 
\beta (t)+c_{2} & -\alpha (t)+c_{1}%
\end{array}%
\right) \,,  \label{Iaf}
\end{equation}%
where 
\begin{eqnarray}
\alpha (t) &=&\alpha _{1}\cos {2\omega t}+\alpha _{2}\sin {2\omega t}+\frac{%
c_{1}}{2}\,,  \nonumber \\
\beta (t) &=&\alpha _{1}\sin {2\omega t}-\alpha _{2}\cos {2\omega t}-\frac{%
c_{2}}{2}\,,  \label{ab}
\end{eqnarray}%
with $\alpha _{1}$, $\alpha _{2}$, $c_{1}$, and $c_{2}$ denoting arbitrary
constants. Therefore, as mentioned before, we can construct the dynamical
invariant such that it is independent of $\gamma _{d}$. The super-operator $%
\mathcal{I}(t)$ given in Eq.~(\ref{Iaf}) has a basis of eigenvectors as long
as $4(\alpha _{1}^{2}+\alpha_2 ^{2})\neq c_{2}^{2}$. This can be adjusted with
no problem since we are free to set the constants. The operator $\mathcal{I}%
(t)$ is in general non-Hermitian, which means that the left and right bases
will not be related by a transpose conjugation operation. The cyclic GPs $%
\varphi _{1}$ and $\varphi _{2}$ associated with the right vectors $|%
\mathcal{D}_{1}\rangle \rangle $ and $|\mathcal{D}_{2}\rangle \rangle $,
respectively, can be computed as given by Eq.~(\ref{cpgp}), yielding 
\begin{eqnarray}
\varphi _{1} &=&-\int_{0}^{t}\langle \langle \mathcal{E}_{1}|\frac{\partial 
}{\partial t^{\prime }}|\mathcal{D}_{1}\rangle \rangle dt^{\prime }\,, \\
\varphi _{2} &=&-\int_{0}^{t}\langle \langle \mathcal{E}_{2}|\frac{\partial 
}{\partial t^{\prime }}|\mathcal{D}_{2}\rangle \rangle dt^{\prime }\,,
\label{gp1}
\end{eqnarray}%
Indeed, by choosing a cyclic path for the basis vectors, we set $t=2\pi
/\omega $. Therefore, we obtain 
\begin{eqnarray}
\varphi _{1} &=&-2\pi \,\frac{c_{2}v_{1}+2v_{2}\sqrt{-(v_{1}/v_{3})^{2}}}{%
v_{1}v_{3}}\,,  \label{phideph1} \nonumber \\
\varphi _{2} &=&-\varphi _{1}\,,
\label{phideph2}
\end{eqnarray}%
where $v_{1}\equiv 2\alpha _{2}+c_{2}$, $v_{2}\equiv \alpha _{1}^{2}+\alpha
_{2}^{2}$, and $v_{3}\equiv \sqrt{4v_{2}-c_{2}^{2}}$. Note that the GP
depends on the particular choice of the super-operator $\mathcal{I}(t)$,
since it depdends on the values of $\alpha _{1}$, $\alpha _{2}$, and $c_{2}$%
. Indeed, different choices of $\mathcal{I}(t)$ will imply in distinct right
and left bases. An interesting particular case is the choice $c_{2}=0$. In
this situation, we obtain 
\begin{eqnarray}
\varphi _{1} &=&-i\pi \frac{\left\vert \alpha _{2}\right\vert }{\alpha _{2}}%
=-i\pi \,{\text{sign}}(\alpha _{2})\,,  \label{gp1f} \\
\varphi _{2} &=&+i\pi \frac{\left\vert \alpha _{2}\right\vert }{\alpha _{2}}%
=+i\pi \,{\text{sign}}(\alpha _{2})\,.  \label{gp2f}
\end{eqnarray}%
Note that, besides robustness against dephasing, the GPs given by Eqs.~(\ref%
{gp1f}) and~(\ref{gp2f}) display only an oscillating (imaginary) term. The
loss of visibility typical in open systems, which is given by the presence of damping real exponentals, is absent for the 
GP in the case $c_{2}=0$. Naturally, a loss of visibility may still come (and indeed it does) from the dynamical phase.


\subsection{Robustness under spontaneous emission}


Now let us analyze the robustness of the GP against spontaneous emission,
which is modelled by $\Gamma =\gamma _{se}\sigma _{-}$, with $\sigma
_{-}=\sigma _{x}-i\sigma _{y}$. In this case, the Lindblad super-operator is
given by (see Appendix~\ref{A}) 
\begin{equation}
\mathcal{L}=\left( 
\begin{array}{cccc}
0 & 0 & 0 & 0 \\ 
0 & -2\gamma _{se}^{2} & -\omega  & 0 \\ 
0 & \omega  & -2\gamma _{se}^{2} & 0 \\ 
4\gamma _{se}^{2} & 0 & 0 & -4\gamma _{se}^{2}%
\end{array}%
\right) \,.  \label{Lse}
\end{equation}%
The super-operator $\mathcal{L}$ motivates the proposal of the dynamical
invariant 
\begin{equation}
\mathcal{I}(t)=\left( 
\begin{array}{cccc}
q(t) & 0 & 0 & p(t) \\ 
0 & \alpha (t) & \beta (t) & 0 \\ 
0 & \gamma (t) & \delta (t) & 0 \\ 
x(t) & 0 & 0 & y(t)%
\end{array}%
\right) \,,  \label{Ise}
\end{equation}%
where the matrix elemtents are arbitrary time-dependent functions. The
commutator $\left[ \mathcal{L},\mathcal{I}\right] $ is now given by 
\begin{equation}
\left[ \mathcal{L},\mathcal{I}\right] =\left( 
\begin{array}{cccc}
4\gamma _{se}^{2}p & 0 & 0 & 4\gamma _{se}^{2}p \\ 
0 & -\varepsilon \,\omega  & \eta \,\omega  & 0 \\ 
0 & \eta \,\omega  & \varepsilon \,\omega  & 0 \\ 
-4\gamma _{se}^{2}(q+x-y) & 0 & 0 & -4\gamma _{se}^{2}p%
\end{array}%
\right) \,,  \label{LIComm2}
\end{equation}%
where $\varepsilon =\beta +\gamma $ and $\eta =\alpha -\delta $. We observe
that the commutator is splitted out in two submatrices. The internal
submatrix is identical to that obtained from dephasing [see Eq.~(\ref{LIComm}%
)], being independent of the decoherence parameter $\gamma _{se}$. In order
to ensure robustness for the external submatrix, we must impose $p=0$ 
(implying from Eq.~(\ref{di}) that both $q$ and $y$ are constants) and  
$q=y-x$ (implying that $x$ is also a constant).
Since, as given by Eq.~(\ref{Ise}), the internal and the external submatrix are decoupled, 
only the internal submatrix will contribute for the GP (the
constant elements of the external submatrix will desappear in the
computation of the GP, due to the time derivative). This means that: (i) the
invariant super-operator $\mathcal{I}(t)$ for spontaneous emission given by
Eq.~(\ref{Ise}) will produce the same GP as that obtained for dephasing;
(ii) since $\mathcal{I}(t)$ can be non-trivially defined as independent of $%
\gamma _{se}$ then the non-adiabatic GP acquired by $\rho $ in the basis of $%
\mathcal{I}(t)$ is robust against spontaneous emission. The robutness of the geometric 
phase under spontaneous emission appears here as a consequence of the expansion of 
the density operator $\rho$ in the basis of a suitably chosen invariant super-operator 
(see, e.g., Ref.~\cite{Carollo:03} for an analysis based on quantum trajectories of a 
geometric phase which is non-robust against spontaneous emission).

\subsection{An example of non-robustness: bit-flip}

\begin{figure}[ht]
\includegraphics[angle=0,scale=0.31]{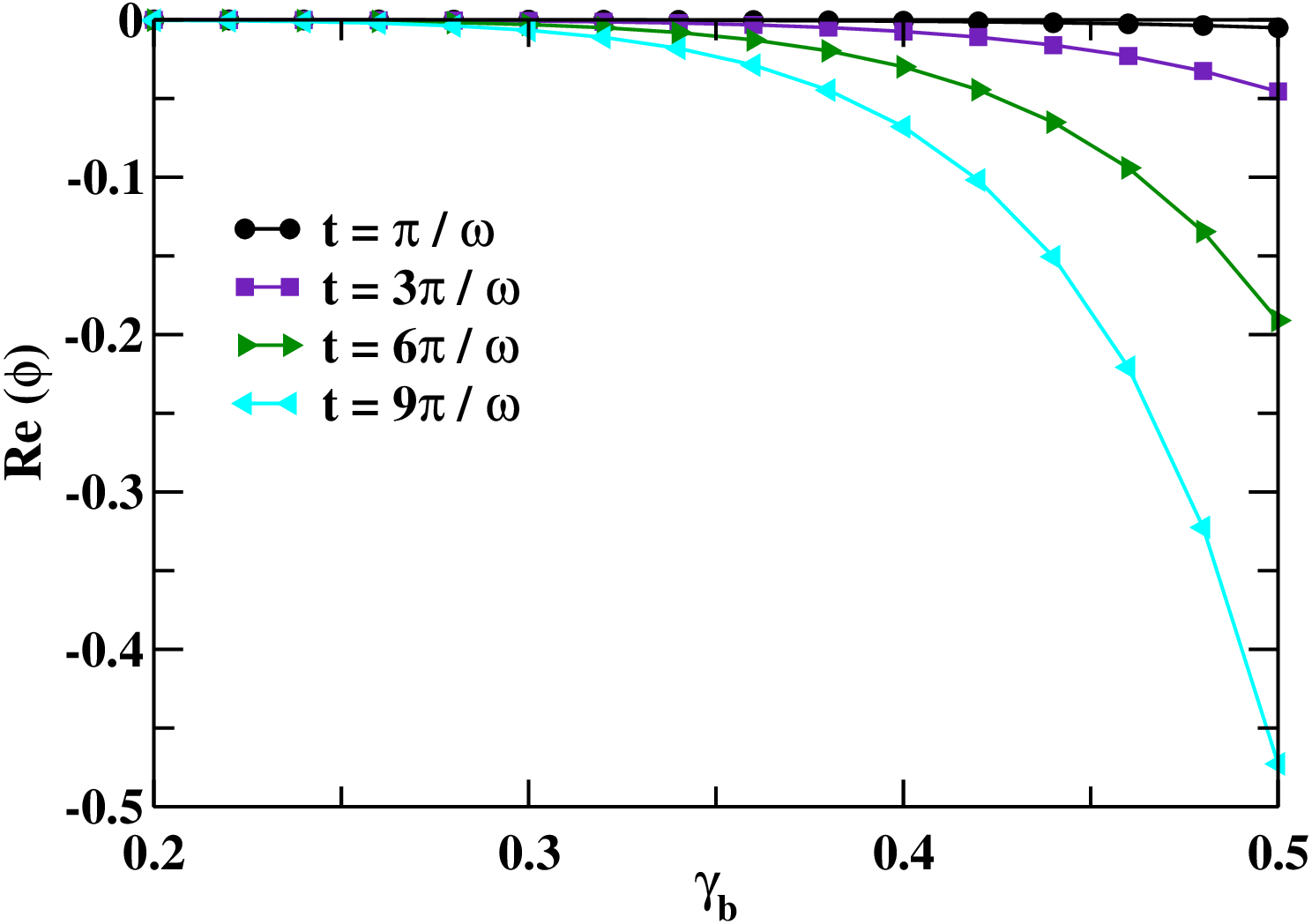}
\caption{(Color online) Real part of the geometric phase for a two-level system under 
bit-flip as a function of the decoherence parameter $\gamma_b$ (in units such that $\omega=1$).}
\label{f1}
\end{figure}

Robustness will not be present for arbitrary processes. For instance,
consider the case of bit-flip, i.e. $\Gamma =\gamma _{b}\sigma _{x}$. In
this case, the Lindblad super-operator reads (see Appendix~\ref{A}) 
\begin{equation}
\mathcal{L}=\left( 
\begin{array}{cccc}
0 & 0 & 0 & 0 \\ 
0 & 0 & -\omega  & 0 \\ 
0 & \omega  & -2\gamma _{b}^{2} & 0 \\ 
0 & 0 & 0 & -2\gamma _{b}^{2}%
\end{array}%
\right) \,.  \label{Lbf}
\end{equation}%
Consider that we propose the dynamical invariant $\mathcal{I}$ given by Eq.~(%
\ref{Ise}). The commutator $\left[ \mathcal{L},\mathcal{I}\right]$ now
yields 
\begin{equation}
\left[ \mathcal{L},\mathcal{I}\right] =\left( 
\begin{array}{cccc}
0 & 0 & 0 & 2\gamma _{b}^{2}p \\ 
0 & -\varepsilon \,\omega  & 2\beta \gamma _{b}^{2}+\eta \,\omega  & 0 \\ 
0 & -2\gamma \gamma _{b}^{2}+\eta \,\omega  & \varepsilon \,\omega  & 0 \\ 
-2\gamma _{b}^{2}x & 0 & 0 & 0%
\end{array}%
\right) \,,  \label{LIComm3}
\end{equation}%
where, as defined for the case of spontaneous emission, $\varepsilon =\beta
+\gamma $ and $\eta =\alpha -\delta $. Therefore, the requirement of
independence of $\gamma _{b}$ yields $x=0$, $p=0$, $\omega (\alpha -\delta
)=-2\beta \gamma _{b}^{2}$, and $\omega (\alpha -\delta )=-2\gamma \gamma
_{b}^{2}$. Then, by using Eqs.~(\ref{ab}), we obtain $\alpha =c_{1}/2$ and $%
\beta =-c_{2}/4$ which, from Eq.~(\ref{Iaf}), imply that $\alpha $, $\beta $%
, $\gamma $, and $\delta $ are constants. Moreover, requiring Eq.~(\ref{di})
for the dynamical invariant, we also find that $q$ and $y$ are constants.
Therefore $\mathcal{I}$ as given by Eq.~(\ref{Ise}) cannot result in
non-vanishing GPs which are robust against bit-flip, since the robust
dynamical invariant obtained is trivially constant. 
Thus, let us turn to the case of time-dependent ${\cal I}(t)$ and  
explicitly analyze the dependence of the geometric phase on the parameter $\gamma_b$. 
By taking $x=0$ and $p=0$ in Eq.~(\ref{LIComm3}), we can choose the dynamical invariant as 
\begin{equation}
\mathcal{I} =\left( 
\begin{array}{cccc}
0 & 0 & 0 & 0 \\ 
0 & \alpha(t) & \beta(t)  & 0 \\ 
0 & \gamma(t) & \delta(t)  & 0 \\ 
0 & 0 & 0 & 0
\end{array}
\right) \, ,  \label{Ibf}
\end{equation}
where now the functions $\alpha(t)$, $\beta(t)$, $\gamma(t)$, and $\delta(t)$ satisfy the 
following set of differential equations
\begin{eqnarray}
\dot{\alpha} &=& - \left( \beta + \gamma \right) \omega \label{bfdiffeqs1} \\
\dot{\beta}  &=& 2\beta\gamma_b^2 + \left( \alpha - \delta \right) \omega  \\
\dot{\gamma} &=& -2\gamma\gamma_b^2 + \left( \alpha - \delta \right) \omega  \\
\dot{\delta} &=&   \left( \beta + \gamma \right) \omega 
\label{bfdiffeqsf}
\end{eqnarray}
The solution of Eqs.~(\ref{bfdiffeqs1})-(\ref{bfdiffeqsf}) can be written as
\begin{eqnarray}
\alpha(t) &=& \omega \frac{\left(-\varepsilon_1 e^{2\xi t} + \varepsilon_2 e^{-2\xi t}\right)}
{2\xi}+\alpha_1 \, ,\nonumber \\
\beta(t) &=& \frac{\varepsilon(t)+\sigma(t)}{2} \, ,\nonumber \\
\gamma(t) &=& \frac{\varepsilon(t)-\sigma(t)}{2} \, , \nonumber \\
\delta(t) &=& -\alpha(t) + c_1 \, ,
\end{eqnarray}
where $\alpha_1$, $\varepsilon_1$, $\varepsilon_2$, and $c_1$ are constants, 
$\xi=(\gamma_b^4-\omega^2)^{1/2}$ and 
\begin{eqnarray}
\varepsilon(t) &=& \varepsilon_1 e^{2\xi t} + \varepsilon_2 e^{-2\xi t} \, ,\nonumber \\
\sigma(t) &=& \gamma_b^2 \left( \frac{\varepsilon_1 e^{2\xi t} - \varepsilon_2 e^{-2\xi t}}{\xi}
\right) + \sigma_1 \, , 
\end{eqnarray}
with $\sigma_1$ satisfying $\gamma_b^2 \sigma_1 + \omega ( 2\alpha_1 - c_1 ) = 0$. 
We are free to set the initial conditions which define the 
dynamical invariant ${\cal I}(t)$. Distinct choices of ${\cal I}(t)$ will imply in different 
GPs acquired by the basis vectors $|{\cal D}_\alpha^{(i)}\rangle\rangle$ that 
expand the density operator. In order to consider a concrete example, we set $\sigma_1=0$, 
which implies $c_1 = 2 \alpha_1$. Moreover, we take $\varepsilon_1= -0.5$, and $\varepsilon_2=1$. 
By adopting these values, we plot in Fig.~\ref{f1} the real part of the GP 
$\phi$, given by Eq.~(\ref{opgp}), as a function of 
the decoherence parameter for several fixed times. 

This GP is non-cyclic and evaluated for the eigenstate of $\cal I$ associated with the eigenvalue $\alpha_1 - \sqrt{\varepsilon_1 \varepsilon_2}$ (the GP is independent of $\alpha_1$). 
Note that the visibility of $\phi$ decreases faster as we increase the evolution time $t$. 
Concerning the imaginary part of $\phi$, it can be shown that it is independent 
of $\gamma_b$ for a given time $t$.

\begin{figure}[ht]
\includegraphics[angle=0,scale=0.32]{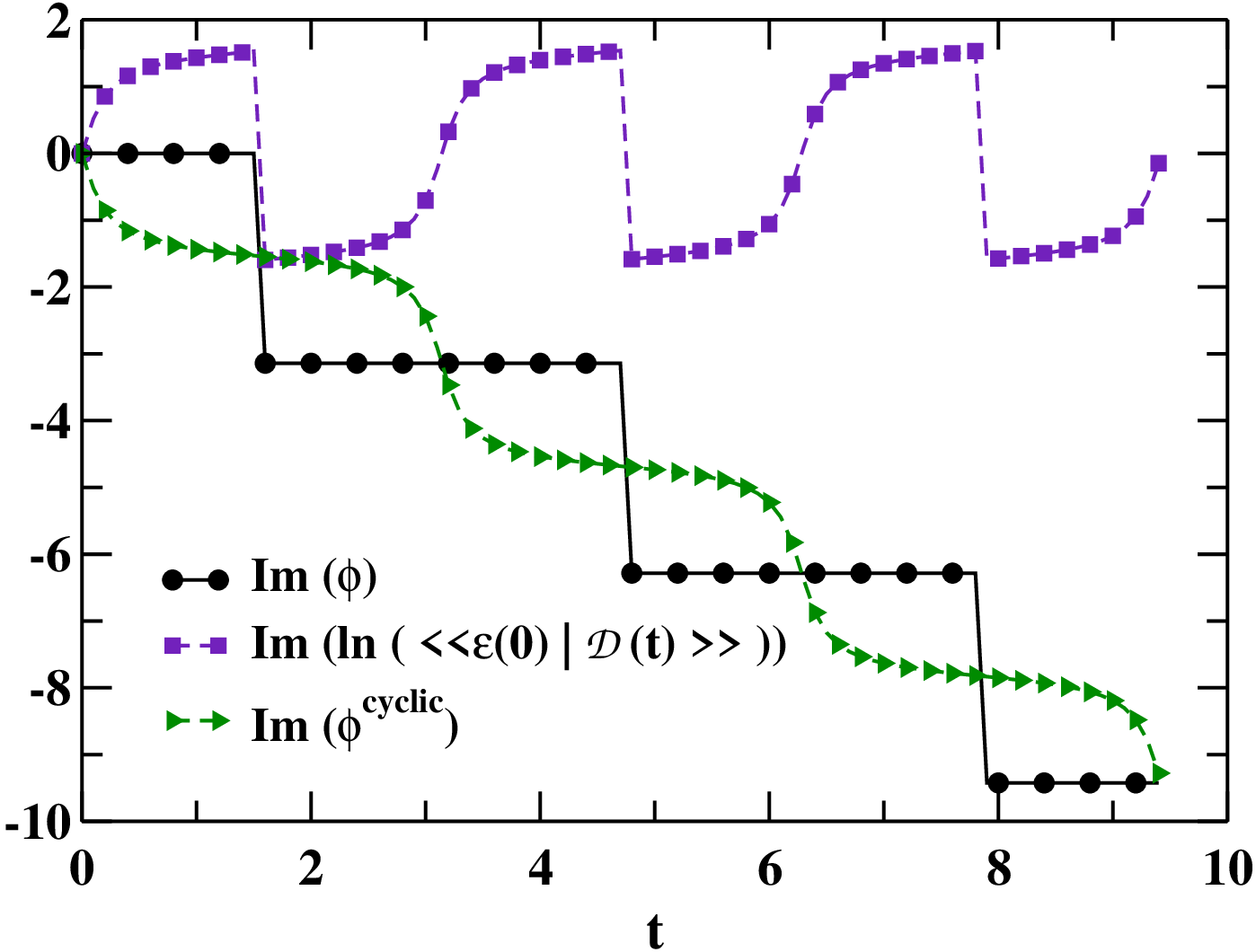}
\caption{(Color online) Imaginary part of the GP for a two-level system under 
bit-flip as a function of time. The decoherence parameter $\gamma_b$ is set to $0.1$ 
(in units such that $\omega=1$).}
\label{f2}
\end{figure}

\begin{figure}[ht]
\includegraphics[angle=0,scale=0.32]{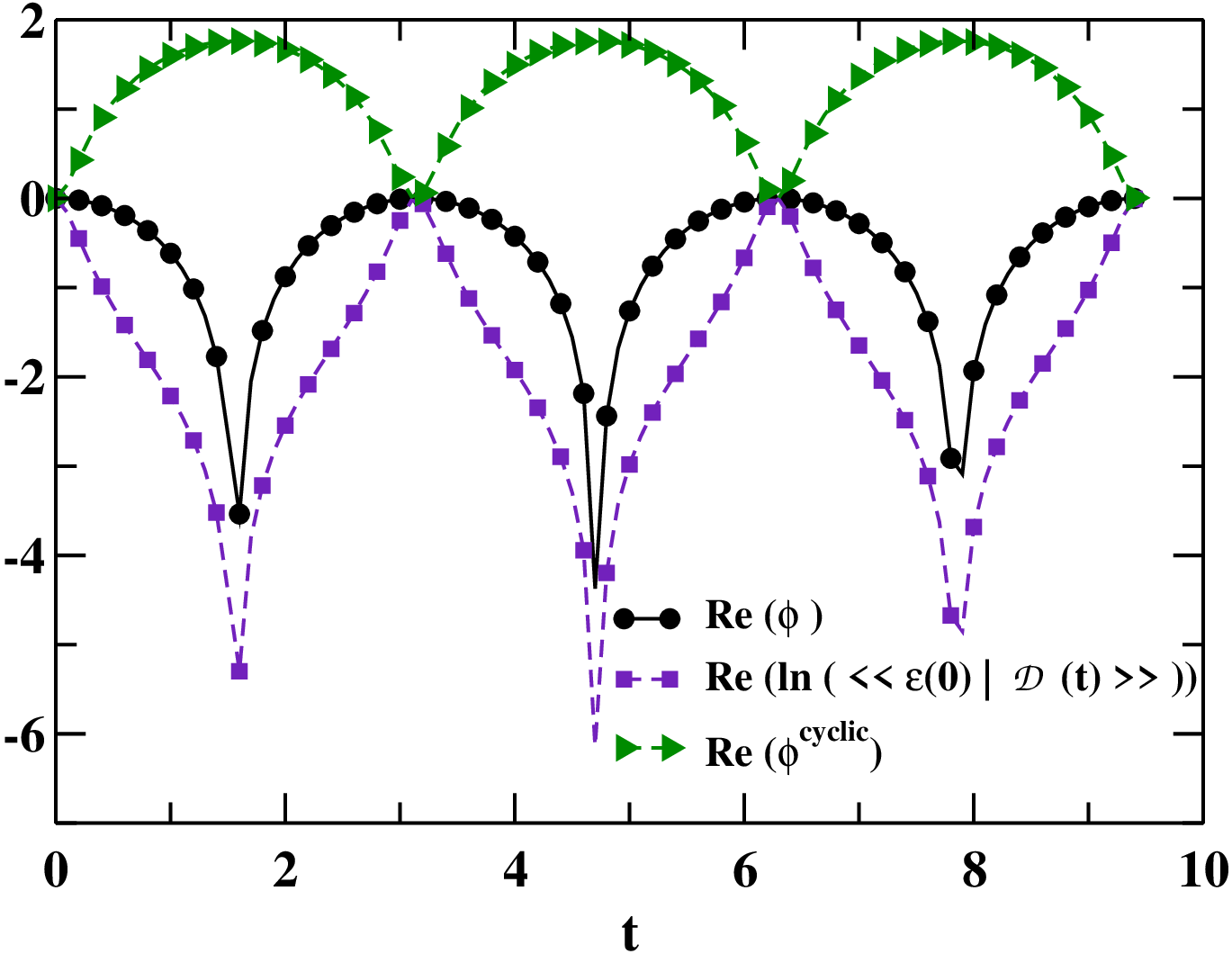}
\caption{(Color online) Real part of the GP for a two-level system under 
bit-flip as a function of time.The decoherence parameter $\gamma_b$ is set to $0.1$ 
(in units such that $\omega=1$).}
\label{f3}
\end{figure}

We can also consider the dependence of the GP as time is varied for a fixed $\gamma_b$. This 
is plotted in Fig.~\ref{f2} and Fig.~\ref{f3}, where we fix $\gamma_b = 0.1$ (in units such 
that $\omega=1$). As we can observe in Fig.~\ref{f2}, the imaginary part of the gauge-invariant GP, which is the sum of $\phi^{\textrm{cyclic}}$ (See Eq.~(\ref{cpgp})) and the logarithmic correction, behaves as a step function of time. The origin of this behavior is the $\ln$ term in Eq.~(\ref{opgp}). Moreover, note that the discontinuities in the imaginary part of the GP are  associated with a pronounced  behavior also in the real part, as exhibited in Fig.~\ref{f3}.   

\subsection{Dynamical phase under decoherence}

Concerning the behavior of the dynamical phase, it will usually not exhibit robustness against decoherence. This is 
due to the fact that the super-operator ${\cal L}$ depends on the decoherence parameters. 
This is in contrast with the invariant super-operator ${\cal I}$, which can be designed to display robustness  
if $\left[{\cal L},{\cal I}\right]$ is independent of the decohering processes (as previously shown for dephasing and 
spontaneous emission). 
Indeed, robustness of the dynamical phase can only be 
achieved whether the integral $\int \langle \langle \mathcal{E}_{\beta}|\mathcal{L}|\mathcal{D}_{\beta}\rangle \rangle dt^{\prime }$ can be 
made independent of decoherence, which turns out to be a non-generic situation. 
As a concrete example, let us consider the dynamical phase for dephasing. In this case, robustness 
is not possible by choosing the invariant operator given in Subsection~\ref{deph}. 
In fact, by explicit computation for a cyclic evolution, we obtain
\begin{eqnarray}
\int_0^{2\pi/\omega}\langle \langle \mathcal{E}_{1}|\mathcal{L}|\mathcal{D}_{1}\rangle \rangle dt^{\prime } &=& 
-\frac{4\pi}{\omega}\gamma_d^2 + \frac{2 c_2 \pi}{v_3} \\
\int_0^{2\pi/\omega}\langle \langle \mathcal{E}_{2}|\mathcal{L}|\mathcal{D}_{2}\rangle \rangle dt^{\prime } &=& 
-\frac{4\pi}{\omega}\gamma_d^2 - \frac{2 c_2 \pi}{v_3}  
\end{eqnarray}
with $v_3$ defined as in Eq.~(\ref{phideph1}). Therefore, notice that no adjust can be done in order to remove the 
dependence of the dynamical phase for an arbitrary $\gamma_d$. As expected, this dependence will induce a damping 
contribution to the visibility of the total phase. 



\section{Conclusions}


We have proposed a generalization of the theory of dynamical invariants to
the context of open quantum systems. This approach can be seen as
an alternative way to solve the master equation, since the construction and 
diagonalization of a dynamical invariant automatically determines the density operator. 
By using this generalization, we have defined in general non-adiabatic GPs acquired
by the density operator during its evolution in Hilbert-Schmidt space.
Moreover, we have delineated a strategy to look for non-adiabatic GPs that
are robust against a given decoherence process. Our method consists in
looking for dynamical invariants such that $\left[ \mathcal{L},\mathcal{I}%
\right] $ is independent of the decohering parameters. As an illustration
of our approach, we have analyzed the GP acquired by a qubit evolving under
decoherence. GP in this case was shown to be robust against both dephasing
and spontaneous emission. Robustness of the non-adiabatic GP against
spontaneous emission is a remarkable feature which may have a positive
impact in geometric quantum computation.
In this direction, a certainly interesting application of our approach is the 
analysis of non-Abelian geometric phases in the tripod-linkage system of atomic 
states~\cite{Unanyan:99,Theuer:99,Duan:01,Faoro:03}. We left this topic for further research.

\begin{acknowledgments}
We gratefully acknowledge financial support from the Brazilian agencies CNPq
and FAPERJ (to M.S.S.), UFABC (to E.I.D.), and CNPq and FAPESP (to M.H.Y.M.).
We also thank Prof. Daniel Lidar for useful comments.
\end{acknowledgments}

\vspace{0.7 cm}

\appendix


\section{Lindblad super-operator for a two-level system under decoherence}


\label{A}

Let us illustrate the construction of the Lindblad super-operator $\mathcal{L%
}$ by examining a two-level system described by the free Hamiltonian given
by Eq.~(\ref{Htls}). We will consider the following decohering process 
\begin{equation}
\Gamma (t)=\alpha _{1}(t)\sigma _{x}+\alpha _{2}(t)\sigma _{y}+\alpha
_{3}(t)\sigma _{z}=\sum_{i=1}^{3}\alpha _{i}(t)\sigma _{i},  \label{Gammasp}
\end{equation}%
where $\sigma _{1}\equiv \sigma _{x}$, $\sigma _{2}\equiv \sigma _{y}$, and $%
\sigma _{3}\equiv \sigma _{z}$. Note that $\Gamma (t)$ describes an
arbitrary single decoherence process for a two-level system. For instance,
for dephasing, we would take $\alpha _{1}=\alpha _{2}=0$. For the density
operator, we can take the expression 
\begin{equation}
\rho (t)=\frac{1}{2}\left( I+\vec{v}\cdot \vec{\sigma}\right) =\frac{1}{2}%
\left( I+v_{1}\sigma _{x}+v_{2}\sigma _{y}+v_{3}\sigma _{z}\right) ,
\label{densop}
\end{equation}%
where $I$ is the two-dimensional identity operator and $\vec{v}$ is the
coherence vector. By inserting Eqs.~(\ref{Htls}),~(\ref{Gammasp}), and (\ref%
{densop}) into the Lindblad equation~(\ref{eq:t-Lind3}) we obtain 
\begin{widetext}
\begin{equation}
\frac{\partial{\rho}}{\partial t} = \frac{\omega}{2} \left( v_1 \sigma_2 - v_2 \sigma_1 \right)+
\sum_{i,j} \frac{\alpha_i^\dagger \alpha_j}{2} \left(v_i \sigma_j + v_j \sigma_i \right) 
- \sum_{i,j} \left| \alpha_i \right|^2 v_j \sigma_j 
- \sum_{i,j,k} i \varepsilon_{ijk} \alpha_i^\dagger \alpha_j \sigma_k 
\label{eq:t-Lindapp}
\end{equation}
\end{widetext}where we have made use of the auxiliary expressions 
\begin{equation}
\sigma _{i}\sigma _{j}=i\varepsilon _{ijk}\sigma _{k}+\delta
_{ij}I\,,\,\,\,\,\,\,\,\varepsilon _{ijk}\varepsilon _{pqk}=\delta
_{ip}\delta _{jq}-\delta _{iq}\delta _{jp},
\end{equation}%
with the repeated indices $k$ summed over and with $\varepsilon _{ijk}$
denoting the Levi-Civita symbol (it is $1$ if $(i,j,k)$ is an even
permutation of $(1,2,3)$, $-1$ if it is an odd permutation, and 0 if any
index is repeated). Factoring out the components in each $\sigma _{i}$%
-direction, Eq.~(\ref{eq:t-Lindapp}) can be rewritten as 
\begin{widetext}
\begin{eqnarray}
\frac{\partial{\rho}}{\partial t} &=&
\left[ -\frac{\omega v_2}{2} + \left(\alpha_1^\dagger \alpha_2 + \alpha_1 \alpha_2^\dagger \right) \frac{v_2}{2} 
+\left(\alpha_1^\dagger \alpha_3 + \alpha_1 \alpha_3^\dagger \right) \frac{v_3}{2} 
-\left( \left| \alpha_2 \right|^2 + \left| \alpha_3 \right|^2 \right) v_1
+ i \left( \alpha_2^\dagger \alpha_3 - \alpha_2 \alpha_3^\dagger \right)\right] \sigma_1 
\nonumber \\
&&+\left[ \frac{\omega v_1}{2} + \left(\alpha_1^\dagger \alpha_2 + \alpha_1 \alpha_2^\dagger \right) \frac{v_1}{2} 
+\left(\alpha_2^\dagger \alpha_3 + \alpha_2 \alpha_3^\dagger \right) \frac{v_3}{2} 
-\left( \left| \alpha_1 \right|^2 + \left| \alpha_3 \right|^2 \right) v_2
- i \left( \alpha_1^\dagger \alpha_3 - \alpha_1 \alpha_3^\dagger \right)\right] \sigma_2 
\nonumber \\
&&+\left[ \left(\alpha_1^\dagger \alpha_3 + \alpha_1 \alpha_3^\dagger \right) \frac{v_1}{2} 
+\left(\alpha_2^\dagger \alpha_3 + \alpha_2 \alpha_3^\dagger \right) \frac{v_2}{2} 
-\left( \left| \alpha_1 \right|^2 + \left| \alpha_2 \right|^2 \right) v_3
+ i \left( \alpha_1^\dagger \alpha_2 - \alpha_1 \alpha_2^\dagger \right)\right] \sigma_3 
\label{eq:t-Lind4}
\end{eqnarray}
\end{widetext}Taking $\rho (t)$ as a vector in Hilbert-Schmidt space and
using Eq.~(\ref{densop}), we can write 
\begin{equation}
|\rho (t)\rangle \rangle =\frac{1}{2}\left( 
\begin{array}{c}
1 \\ 
v_{1} \\ 
v_{2} \\ 
v_{3}%
\end{array}%
\right)  \label{rhovec}
\end{equation}%
where $|\rho (t)\rangle \rangle $ is expressed in the basis $\{I,\sigma
_{1},\sigma _{2},\sigma _{3}\}$. Therefore, by inserting Eq.~(\ref{rhovec})
and Eq.~(\ref{eq:t-Lind4}) (for $\frac{\partial {\rho }}{\partial t}$) into
Eq.~(\ref{eq:t-Lind2}), we obtain the Lindblad super-operator $\mathcal{L}$ 
\begin{widetext}
\begin{equation}
{\cal L} = \left(
\begin{array}{cccc}
0 & 0 & 0 & 0 \\
2i\left( \alpha_2^\dagger \alpha_3 - \alpha_2 \alpha_3^\dagger \right) & 
-2\left( \left| \alpha_2 \right|^2 + \left| \alpha_3 \right|^2 \right) & 
-\omega + \left(\alpha_1^\dagger \alpha_2 + \alpha_1 \alpha_2^\dagger \right) & 
\left(\alpha_1^\dagger \alpha_3 + \alpha_1 \alpha_3^\dagger \right) \\
-2i\left( \alpha_1^\dagger \alpha_3 - \alpha_1 \alpha_3^\dagger \right) & 
\omega + \left(\alpha_1^\dagger \alpha_2 + \alpha_1 \alpha_2^\dagger \right) &
-2\left( \left| \alpha_1 \right|^2 + \left| \alpha_3 \right|^2 \right) &
\left(\alpha_2^\dagger \alpha_3 + \alpha_2 \alpha_3^\dagger \right) \\ 
2i \left( \alpha_1^\dagger \alpha_2 - \alpha_1 \alpha_2^\dagger \right) &
\left(\alpha_1^\dagger \alpha_3 + \alpha_1 \alpha_3^\dagger \right) & 
\left(\alpha_2^\dagger \alpha_3 + \alpha_2 \alpha_3^\dagger \right) &
-2\left( \left| \alpha_1 \right|^2 + \left| \alpha_2 \right|^2 \right)  
\end{array}
\right)
\label{Lfinal}
\end{equation}
\end{widetext}Some interesting particular cases of Eq.~(\ref{Lfinal}) can be
obtained. For instance, for dephasing, we have $\alpha _{1}=\alpha _{2}=0$
and $\alpha _{3}\equiv \gamma _{d}$, resulting in Eq.~(\ref{Ldp}). Note that
the first column of $\mathcal{L}$ vanishes for dephasing. In fact, this will
be the case whenever the parameters $\alpha _{i}$ are real. An interesting
case of complex $\alpha _{i}$ is given by spontaneous emission, where $%
\alpha _{1}\equiv \gamma $, $\alpha _{2}\equiv -i\gamma $, and $\alpha
_{3}=0 $. In this case, we obtain the super-operator $\mathcal{L}$ shown in
Eq.~(\ref{Lse}).

\end{document}